\documentclass[prd,showpacs,floatfix,twocolumn,,amsmath,amssymb,floatfix]{revtex4}
\usepackage{graphicx,color,dcolumn,booktabs,bm}
\usepackage{longtable,lscape}
\usepackage{txfonts}
\usepackage{overpic}
\usepackage{amssymb}
\usepackage{indentfirst}
\usepackage{epsfig}
\usepackage{epstopdf}   %{feynmp}
\usepackage{slashed}  %for Feynman symbols
\usepackage{cases}
\usepackage{color}
\usepackage{multirow}
\usepackage{graphicx,color,dcolumn,booktabs,bm}

\graphicspath{{Figures/}} %
\usepackage[colorlinks, citecolor=blue,anchorcolor=red,menucolor=red, linkcolor=red,filecolor=red,runcolor=red,urlcolor=blue,frenchlinks=red]{hyperref}

\begin{document}

\title{A hidden-charm pentaquark state in $\Lambda^0_b \to J/\psi p \pi^-$ decay}

\author{En Wang$^{1,5}$}
\author{Hua-Xing Chen$^2$}
\email{hxchen@buaa.edu.cn}
\author{Li-Sheng Geng$^{2,3}$}
\email{lisheng.geng@buaa.edu.cn}
\author{De-Min Li$^{1}$}
\email{lidm@zzu.edu.cn}
\author{Eulogio Oset$^{4,5}$}
\email{Eulogio.Oset@ific.uv.es}
\affiliation{
$^1$Department of Physics, Zhengzhou University, Zhengzhou, Henan 450001, China \\
$^2$School of Physics and Nuclear Energy Engineering and International Research Center for Nuclei and Particles in the Cosmos, Beihang University, Beijing 100191, China \\
$^3$State Key Laboratory of Theoretical Physics, Institute of Theoretical Physics, Chinese Academy of Sciences, Beijing 100190, China \\
$^4$Institute of Modern Physics, Chinese Academy of Sciences, Lanzhou 730000, China \\
$^5$Departamento de F\'{\i}sica Te\'orica and IFIC, Centro Mixto Universidad de Valencia-CSIC Institutos de Investigaci\'on de Paterna, Aptdo. 22085, 46071 Valencia, Spain}

\begin{abstract}
We study here the  $\Lambda_b^0 \to J/\psi p \pi^-$ reaction in analogy to the  $\Lambda^0_b \to J/\psi p K^-$ one, and we note that in both decays there is a sharp structure (dip or peak) in the $J/\psi p$ mass distribution around $4450$ MeV, which is associated in the  $\Lambda^0_b \to J/\psi p K^-$ experiment to an exotic pentaquark baryonic state, although in $\Lambda_b^0 \to J/\psi p \pi^-$ it shows up with relatively low statistics. We analyze the $\Lambda^0_b \to J/\psi p \pi^-$ interaction along the same lines as the  $\Lambda^0_b \to J/\psi p K^-$  one, with the main difference stemming from the reduced Cabibbo strength in the former and the consideration of the $\pi^-p$ final state interaction instead of the $K^-p$ one. We find that with a minimal input, introducing the $\pi^-p$ and $J/\psi p$ interaction in $S$-wave with realistic interactions, and the empirical $P$-wave and  $D$-wave contributions, one can accomplish a qualitative description of the $\pi^-p$ and $J/\psi p$ mass distributions. More importantly, the peak structure followed by a dip of the experimental $J/\psi p$ mass distribution is reproduced with the same input as used to describe the data of $\Lambda^0_b\rightarrow J/\psi p K^-$ reaction. The repercussion for the triangular singularity mechanism, invoked in some works to explain the pentaquark peak, is discussed.
\end{abstract}

\date{\today}
%\keywords{}
\maketitle

\section{Introduction}
\label{sec:intro}

Via the $\Lambda_b^0 \to J/\psi p K^-$ decay, the LHCb Collaboration observed two hidden-charm pentaquark states, $P_c(4380)$ and $P_c(4450)$, in the $J/\psi p$ invariant mass spectrum~\cite{Aaij:2015tga}.
Additionally, their resonance parameters are measured to be
\begin{eqnarray}
\nonumber && M_{P_c(4380)}=4380\pm 8\pm 29\, \rm{MeV} \, ,
\\ && \Gamma_{P_c(4380)}=205\pm18\pm86\, \rm{MeV} \, ,
\\ \nonumber && M_{P_c(4450)}=4449.8\pm 1.7\pm 2.5 \, \rm{MeV} \, ,
\\ \nonumber && \Gamma_{P_c(4450)}=39\pm5\pm19\, \rm{MeV} \, .
\end{eqnarray}
Later, the LHCb Collaboration continued their studies and measured the branching fraction of $\Lambda^0_b \to J/\psi p K^-$~\cite{Aaij:2015fea}
\begin{eqnarray}
&&{\cal B}(\Lambda^0_b \to J/\psi p K^-)\nonumber \\  &=& ( 3.04 \pm 0.04 \pm 0.06  \pm 0.33 {^{+0.43}_{-0.27}} ) \times 10^{-4} \, .
\label{brK}
\end{eqnarray}
At the same time, they also updated the branching fraction of $\Lambda^0_b \to J/\psi p \pi^-$~\cite{Aaij:2015fea}
\begin{eqnarray}
&& {\cal B}(\Lambda^0_b \to J/\psi p \pi^-)\nonumber \\  &=& ( 2.51 \pm 0.08 \pm 0.13 {^{+0.45}_{-0.35}} ) \times 10^{-5} \, ,
\label{brpi}
\end{eqnarray}
which is Cabibbo suppressed compared to the previous one. However, the effect of $P_c(4380)$ and $P_c(4450)$ can still be significant, because the reason for the suppression is the presence of the different Cabibbo-Kobayashi-Maskawa (CKM) matrix elements, $V_{cb}V^*_{cd}$ for $\Lambda^0_b \to J/\psi p \pi^-$ and  $V_{cb}V^*_{cs}$ for $\Lambda^0_b \to J/\psi p K^-$, which are global factors.

In Ref.~\cite{Aaij:2014zoa}, the LHCb collaboration first reported this Cabibbo-suppressed decay $\Lambda^0_b\to J/\psi p \pi^-$.  In Fig.2(d) of Ref.~\cite{Aaij:2014zoa}, there is a peak in the $J/\psi p$ distribution, compatible with the one seen in the decay of $\Lambda^0_B\to J/\psi p K^-$, although unfortunately with relatively low statistics.
Hence, if more data on the $\Lambda^0_b \to J/\psi p \pi^-$ decay are collected by the LHCb collaboration, the $P_c(4380)$ and $P_c(4450)$ may be more clearly seen. This decay
can be a very efficient way for the LHCb to check their previous results. Moreover, the $P_c(4380)$ and $P_c(4450)$
may contribute differently to the $\Lambda_b^0 \to J/\psi p K^-$ and $\Lambda^0_b \to J/\psi p \pi^-$ decays, therefore a careful study of the second decay is  of tremendous value to better understand the properties of the two pentaquark states.  As remarked in Ref.~\cite{Burns:2015dwa}, no claim of an exotic state  was done in Ref.~\cite{Aaij:2014zoa}, and also no reference to the peak of Ref.~\cite{Aaij:2014zoa} was done in Ref.~\cite{Aaij:2015tga}. Yet, the large impact of the work of Ref.~\cite{Aaij:2015tga} is bringing new attention to the $\Lambda_b^0 \to J/\psi p \pi^-$ reaction, which is under reconsideration by the LHCb collaboration~\cite{friends}.

This triggers us to study the $\Lambda^0_b \to J/\psi p \pi^-$ decay, which was also discussed in Refs.~\cite{Burns:2015dwa,Hsiao:2015nna,Cheng:2015cca},
and the following branching fraction was given in Ref.~\cite{Hsiao:2015nna}
\begin{eqnarray}
{{\cal B}(\Lambda^0_b \to P_c \pi^-) {\cal B}(P_c \to J/\psi p) \over {\cal B}(\Lambda^0_b \to P_c K^-) {\cal B}(P_c \to J/\psi p)} = 0.58 \pm 0.05 \, ,
\end{eqnarray}
by assuming that the productions of both $P_c(4380)$ and $P_c(4450)$ are mainly from the charmless $\Lambda_b$ decays through $b \to \bar u u s$, while their $c\bar c$ contents are from the intrinsic charms in the $\Lambda_b$ baryon, i.e. $\Lambda_b[b u d] \to [(\bar u u s) u d] [c\bar c] \to K^- P_c$.

The $\Lambda^0_b \to J/\psi p \pi^-$ decay offers a new possibility  to study $P_c(4380)$ and $P_c(4450)$. Before their observation, the existence of hidden charm pentaquark states had been discussed in Refs.~\cite{Wu:2010jy,Yang:2011wz,Xiao:2013yca,Uchino:2015uha,Karliner:2015ina,Garzon:2015zva,Wang:2011rga,Yuan:2012wz,Huang:2013mua} using various methods. Yet, the experimental observation of these states triggered more theoretical works. Various pictures have been proposed for the nature of these states,  such as meson-baryon molecules~\cite{Chen:2015loa,Roca:2015dva,He:2015cea,Huang:2015uda,Meissner:2015mza,Xiao:2015fia,Chen:2015moa}, diquark-diquark-antiquark pentaquarks~\cite{Maiani:2015vwa,Anisovich:2015cia,Ghosh:2015ksa,Wang:2015epa}, compact diquark-triquark pentaquarks~\cite{Lebed:2015tna,Zhu:2015bba},  $\bar{D}$-soliton states~\cite{Scoccola:2015nia}, genuine multiquark states other than molecules~\cite{Mironov:2015ica}, and kinematical effects related to the so-called triangle singularity~\cite{Guo:2015umn,Liu:2015fea,Mikhasenko:2015vca} (for a more extensive summary, see Ref.~\cite{Burns:2015dwa}). In addition to the mass spectrum, the production of these pentaquark states  in various decays and reactions has been studied,  including weak decays of bottom baryons~\cite{Li:2015gta,Cheng:2015cca}, photo-production~\cite{Wang:2015jsa,Kubarovsky:2015aaa,Karliner:2015voa}, and the $\pi^- p \to J/\psi n$ reaction~\cite{Lu:2015fva}. In a recent work~\cite{Wang:2015qlf}, the strong decays of these states have been studied in the molecule picture .

In this work, we shall follow the same approach used in Ref.~\cite{Roca:2015dva} studying
the $\Lambda^0_b \to J/\psi p K^-$ decay. By using the model of the $ \pi^-p$ scattering studied in Ref.~\cite{Inoue:2001ip}
and the data of the $J/\psi p$ scattering studied in Ref.~\cite{Roca:2015dva}, we can fix nearly all the parameters involved in this process,
except one overall strength and two parameters describing the $P$-wave and $D$-wave $ \pi^-p$ scattering, which, however, do not affect the peak of $J/\psi p$.

This paper is organized as follows. In Sec.~\ref{sec:formalism}, we study the weak decay process, $\Lambda^0_b \to J/\psi p \pi^-$, and separate it into three steps, weak decay, hadronization, and final state interactions. This formalism is used to perform numerical analyses in Sec.~\ref{sec:results}, and some concluding remarks are given in Sec.~\ref{sec:summary}.

\section{Formalism}
\label{sec:formalism}

\begin{figure}[!htb]
\centering
\includegraphics[width=0.35\textwidth]{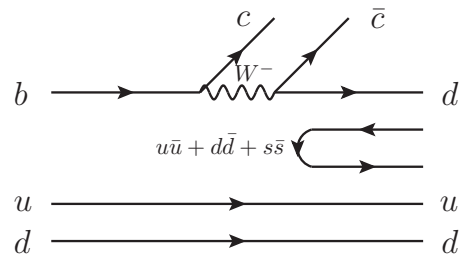}
\caption{Weak decay and hadronization mechanism of $\Lambda^0_b \to J/\psi p \pi^-$ decay.}
\label{fig:weakdecay}
\end{figure}

In this section, we present the formalism of the $\Lambda^0_b \to J/\psi p \pi^-$ decay, as depicted in Fig.~\ref{fig:weakdecay}. This process is suppressed compared to $\Lambda_b^0\rightarrow J/\psi p K^-$, but the effect of pentaquark states can still be important, as discussed in the Introduction, and we will see explicitly in the following.
We shall follow the formalism proposed in Ref.~\cite{Liang:2014tia}, which has been applied to study the decays of $\Lambda_b^0\rightarrow J/\psi p K^-$~\cite{Roca:2015dva,Roca:2015tea}, $\Xi_b^- \rightarrow J/\psi \Lambda K^-$~\cite{Chen:2015sxa} and $\Lambda_b \rightarrow J/\psi K \Xi$~\cite{Feijoo:2015cca}.
In Ref.~\cite{Roca:2015dva}, only the narrow peak of the $J/\psi p$ distribution, associated to the $P_c(4450)$ state, was interpreted as a molecular state of $\bar{D}^*\Sigma_c$ type decaying into $J/\psi p$. Following Ref.~\cite{Liang:2014tia}, where $B^0$ and $B^0_s$ decays into $J/\psi \pi^+\pi^-$ were studied,  the $\Lambda^0_b \to J/\psi K^- p$ decay was addressed in Ref.~\cite{Roca:2015tea}. The $\Lambda^0_b$ decay can be separated into three steps, weak decay, hadronization, and final state interaction. We shall discuss them in the following subsections.

In the experimental paper~\cite{Aaij:2014zoa}, there are some structures in the $\pi^-p$ mass distribution, which are associated to the contributions of the $N^*(1440)$ $(1/2^+)$, $N^*(1535)$ $(1/2^-)$, $N^*(1650)$ $(1/2^-)$, and $N^*(1520)$ $(3/2^-)$ resonances, although a partial wave analysis was not done there. In our picture, we must keep in mind that the $P_c(4450)$ state was associated in Ref.~\cite{Roca:2015dva} to a molecular state of $\bar{D}^*\Sigma_c$ interacting in $S$-wave, which decays into $J/\psi p$ also in $S$-wave. Our formalism contains this $J/\psi p$ interaction, and for consistency we must also take into account the $\pi^-p$ $S$-wave interaction, to allow for possible interference. For this purpose, we shall use the chiral unitary approach for $\pi^-p$ and coupled channels developed in Ref.~\cite{Inoue:2001ip}. This formalism, considering only pseudoscalar-baryon interaction, produced the $N^*(1535)$, although using somewhat unnatural subtraction constants. The $N^*(1650)$ did not show up in this approach, but it did in the related one of Ref.~\cite{Nieves:2001wt}, which relied upon off shell extrapolation of the amplitudes. Within the on-shell factorization approach of Ref.~\cite{Inoue:2001ip}, the $N^*(1650)$ was recovered in Ref.~\cite{Garzon:2012np} by including $\rho N$ and $\pi \Delta$ extra channels. The experimental data of Ref.~\cite{Aaij:2014zoa} shows a moderate peak for the $N^*(1650)$ and a more pronounced peak for the $N^*(1535)$, so the use of the approach of Ref.~\citep{Inoue:2001ip} is sufficient for our purpose.

On the other hand, the Roper $N^*(1440)$ appears in $P$-wave, and the $N^*(1520)$, dynamically produced from the interaction of pseudoscalar-baryon decuplet~\cite{Sarkar:2004jh}, appears in $D$-wave of the $\pi^-p$ system. Both of them  would contribute incoherently to the $\pi^-p$ mass distribution. We shall take into account some contributions of $P$-wave and $D$-wave to show that they do not modify the the $J/\psi p$ mass distribution around the peak.

\subsection{Weak decay and hadronization}

The quark content of $\Lambda^0_b$ is $bud$, where the $u$ and $d$ quarks are in a state of spin zero and isospin zero. Hence, these two light quarks are flavor antisymmetric, and the following simplified notation
can be used to describe the $\Lambda^0_b$:
\begin{equation}
\Lambda^0_b = |b\rangle |ud\rangle \Rightarrow \frac{1}{\sqrt{2}} |b\rangle (|u\rangle|d\rangle - |d\rangle|u\rangle) = \frac{1}{\sqrt{2}}b(ud-du) \, .
\end{equation}
In the $\Lambda^0_b$ decay, the $b$ quark first decays into a $c$ quark by emitting a $W^-$ boson, then the $W^-$ decays into a pair of $\bar{c}$ and $d$ quarks, depicted in Fig.~\ref{fig:weakdecay}, which gives an overall suppressed factor, $V_{cb}V^*_{cd}$:
\begin{equation}
\Lambda^0_b \Rightarrow [V_{cb}]c W^- \frac{1}{\sqrt{2}}(ud-du) \Rightarrow [V_{cb}V^*_{cd}]  c \bar c d\frac{1}{\sqrt{2}}(ud-du) \, .
\end{equation}
As in Refs.~\cite{Feijoo:2015cca,Roca:2015tea}, we will assume that the initial $ud$ pair of the $\Lambda^0_b$ acts as a spectator, and is transferred to the final baryon. This is suggested by the analysis of the data of Ref.~\cite{Aaij:2015tga}, where in the $K^-p$ mass distribution only $\Lambda^*$ states contributed, naturally coming from a final $s$ quark plus the $ud$ pair in $I=0$, acting as a spectator in the process. In the present case, the $s$ final quark in the $\Lambda^0_b\rightarrow J/\psi sud$ is replaced by a $d$ quark, and the $ud$ pair still remains as an $I=0$ state.

In order to have a $\pi^-p$ at the end, the $d$ quark and  $ud$ pair must hadronize, which is accomplished by introducing an extra $\bar{q}q$ pair with the quantum numbers of the vacuum, $\bar{u}u+\bar{d}d+\bar{s}s$. This is introduced between two quarks and it is clear that the $d$ quark coming from the $b$ decay must enter the hadronization process. This is so because the $\pi^-p$ interaction in $S$-wave has negative parity. Since the $ud$ spectator pair has positive parity, it is the weak-decay $d$ quark that must carry negative parity prior to the hadronization and be in an $L=1$ state, and it must be this physical process which brings the quark back to its ground state in the final $\pi^-p$ hadronic state. In order to keep the original $ud$ quark as a spectator  and be transferred to the final baryon, the $d$ quark from the $b$ decay must go to the outgoing pion. Then the hadronization process  proceeds as
\begin{equation}
\Lambda^0_b \Rightarrow J/\psi d (\bar u u + \bar d d + \bar s s) \frac{1}{\sqrt{2}}(ud-du) \, ,
\end{equation}
where the first $d$ quark together with the next $\bar{q}$ will produce a meson and the remaining three quarks a baryon.
\begin{figure}[!htb]
\centering
\includegraphics[width=0.35\textwidth]{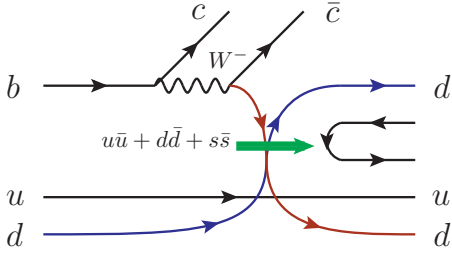}
\caption{Weak decay and hadronization mechanism of $\Lambda^0_b \to J/\psi p \pi^-$ decay, where the $d$ quark contained in $\Lambda^0_b$ forms the meson together with the anti-quark from the vacuum and the rest three quarks form the baryon.}
\label{fig:weakdecay2}
\end{figure}

In principle, the $d$ quark contained in $\Lambda^0_b$ could also form the meson together with the anti-quark from the vacuum and the remaining three quarks form the baryon, as depicted in Fig.~\ref{fig:weakdecay2}. However, this diagram is much suppressed due to the large momentum transferred to the original $d$ quark. Further arguments are given in Ref.~\cite{Miyahara:2015cja}. There is also an interesting experimental feature supporting the spectator assumption for the original $ud$ pair. In this case, the final state has only the isospin of the $d$ quark coming from the $b$ decay and hence the total isospin of the final state is 1/2. This is supported by the experiment of Ref.~\cite{Aaij:2014zoa}, where in the $\pi^-p$ spectrum there is no trace of $\Delta(1232)$ excitation.

\begin{figure}[!htb]
\centering
\includegraphics[width=0.45\textwidth]{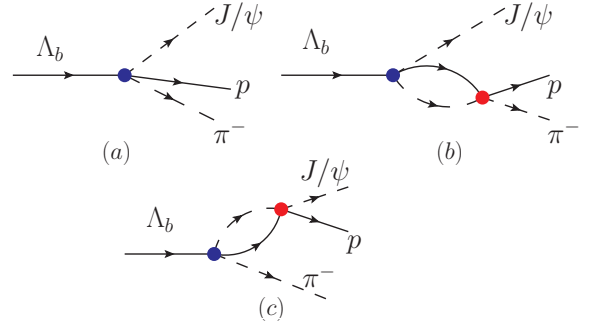}
\caption{(Color online) Diagrams for the $\Lambda^0_b \to J/\psi p \pi^-$ decay: (a) direct $J/\psi p\pi^-$ vertex at tree level, (b) final state interaction of $\pi^- p$, and (c) final state interaction of $J/\psi p$.}
\label{fig:decaydiagram}
\end{figure}

Following the procedure used in Refs.~\cite{Liang:2014tia,Roca:2015tea,Close:1979bt}, we can hadronize the quark combinations $Q=d (\bar{u}u + \bar{d}d + \bar{s}s) \frac{1}{\sqrt{2}}(ud-du)$
into pairs of ground state mesons and baryons, and we find,
%~\footnote{The $u$ and $d$ quarks from the initial $\Lambda^0_b$ couple to isospin $I=0$, this is supported by experiment since one does not see a $\Delta(1232)$ peak in the $\pi^-p$ distribution.}
\begin{equation}
Q \Rightarrow \pi^- p - {1 \over \sqrt2} \pi^0 n + {1 \over \sqrt3} \eta n + \sqrt{\frac{2}{3}} K^0 \Lambda \, ,\label{eq:hadron}
\end{equation}
and hence
\begin{equation}
\label{eq:hadronization}
\Lambda^0_b \Rightarrow J/\psi \times \left[ \pi^- p - {1 \over \sqrt2} \pi^0 n + {1 \over \sqrt3} \eta n + \sqrt{\frac{2}{3}} K^0 \Lambda \right] \, .
\end{equation}
We can clearly see that $J/\psi p \pi^-$ is one of the possible final states, but we still need to consider the final state interaction of the $\pi^- p$  and the other pairs of ground state mesons and baryons, which have been studied in Ref.~\cite{Inoue:2001ip} and known to be very strong.

\subsection{Final state interactions }
\label{sec:fsi}

In this subsection we take into account the final state interaction of ground state mesons and baryons of the octet, as depicted in Fig.~\ref{fig:decaydiagram}(b).
Here we only consider the $S$-wave scattering, and the $P$-wave scattering will be discussed in the next subsection.

%Here we should only care about the total angular momentum, but not parity, which is not conservation any more. The process that goes to the final state of $L=0$, which should dominate the reaction of  $\Lambda^0_b \to J/\psi p %\pi^-$. In this case, we will concentrate in the contribution from all the possible $L=0$, which means $L=0$ between $\pi^-$ and $P$  system and also $J/\psi$ and $\pi^-p$ system. As a result, we would get a contribution from %background and also from the $N^*(1535)$ and $N^*(1650)$.
%In addition, we could also have $L=1$ in $\pi^-p$ system, which will be discussed in the end of this section.

\begin{center}
\begin{table}[ht]
\caption{The coefficients $C_{ij} = C_{ji}$ for the $S$-wave meson baryon scattering in the strangeness $S=0$ and isospin $I=1/2$ sector~\cite{Inoue:2001ip}.}
\vspace{0.5cm}
\begin{tabular}{c|cccccc}\hline \hline
& $K^+ \Sigma^-$ & $K^0 \Sigma^0$ & $K^0 \Lambda$ & $\pi^- p$ & $\pi^0 n$ & $\eta n$
\\ \hline
$K^+ \Sigma^-$ & $1$ & $-\sqrt2$ & 0 & 0 & $-1/\sqrt2$ & $-\sqrt{3/2}$
\\
$K^0 \Sigma^0$ &     & 0 & 0 & $-1/\sqrt2$ & $-1/2$ & $\sqrt3/2$
\\
$K^0 \Lambda$  &     &   & 0 & $-\sqrt{3/2}$ & $\sqrt3/2$ & $-3/2$
\\
$\pi^- p$      &     &   &   & 1 & $-\sqrt2$ & 0
\\
$\pi^0 n$      &     &   &   &   & 0 & 0
\\
$\eta n$       &     &   &   &   &   & 0
\\ \hline \hline
\end{tabular}
\label{tab:cij}
\end{table}
\end{center}

In Ref.~\cite{Inoue:2001ip}, the $S$-wave meson baryon scattering was studied in the strangeness $S = 0$ sector in the coupled channel unitary approach with six channels. The $6\times 6$ $t_{i\,j}$ matrix is given by the Bethe-Salpeter equation,
\begin{equation}
t_{i\,j} = V_{i\,j} + \sum_k V_{i\,k} G_{k} t_{k\,j} \, ,
\end{equation}
where $i,j,k = \left( K^+ \Sigma^-, \, K^0 \Sigma^0, \, K^0 \Lambda, \, \pi^- p, \, \pi^0 n, \, \eta n \right)$.
%Based on these matrices, Eq.~(\ref{eq:hadronization}) can
%be further simplified to be
%\begin{equation}
%\Lambda^0_b \Rightarrow J/\psi \times \sum_i h_i \times i \, ,
%\end{equation}
%where $h_i$ is the weight of the transition:
%\begin{eqnarray}
%&
%h_{K^+ \Sigma^-} = 0,~~
%h_{K^0 \Sigma^0} = 0,~~
%h_{K^0 \Lambda} = \sqrt{\frac{2}{3}},
%&
%\\ \nonumber &h_{\pi^- p}= 1,~~
%h_{\pi^0 n}= - {1 \over \sqrt2},~~
%h_{\eta n}= {1 \over \sqrt3}.
%&
%\end{eqnarray}
%
The matrices $V_{i\,j}$ and $G_{i}$ have both been evaluated in Ref.~\cite{Inoue:2001ip}, and we briefly show the results here. The matrix $V_{i,j}$ is the transition potential obtained from the lowest order meson baryon chiral Lagrangian
\begin{eqnarray}
V_{ij}(s) &=& -C_{ij} \frac{1}{4f_if_j} (2\sqrt{s}-M_i-M_j)
\nonumber \\ && \times \left(\frac{M_i+E_i}{2M_i}\right)^{1/2} \left(\frac{M_j+E_j}{2M_j}\right)^{1/2},
\end{eqnarray}
where $E_i$ and $M_i$ are the energy and mass of the baryon in channel $i$, and the coefficients $C_{ij}$ are shown in Table~\ref{tab:cij}, reflecting the SU(3) flavor symmetry. The couplings $f_i$ are the pseudoscalar decay constants for the $i$ channel, for which we use
\begin{eqnarray}
f_\pi=93~{\rm MeV},~~f_K=1.22 f_\pi,~~f_\eta=1.3 f_\pi.
\end{eqnarray}

The matrix $G_{i}$ is the $G$-function representing the loop integral of a meson and a baryon, for which we adopt the dimensional regularization,
\begin{eqnarray}
G_{i}(s)
&=& i 2 M_i \int \frac{d^4 q}{(2 \pi)^4} \, \frac{1}{(P-q)^2 - M_i^2 + i \epsilon} \, \frac{1}{q^2 - m^2_i + i \epsilon}
\nonumber \\ &=& \frac{2 M_i}{16 \pi^2} \left\{ a_i(\mu) + \ln \frac{M_i^2}{\mu^2} + \frac{m_i^2-M_i^2 + s}{2s} \ln \frac{m_i^2}{M_i^2} + \right.
\nonumber \\ && \phantom{\frac{2 M}{16 \pi^2}} + \frac{\bar{q}_i}{\sqrt{s}} \left[ \ln(s-(M_i^2-m_i^2)+2\bar{q}_i\sqrt{s})\right.
\nonumber \\ && \left. \phantom{\frac{2 M}{16 \pi^2} + \frac{\bar{q}_i}{\sqrt{s}}} \hspace*{-0.3cm} + \ln(s+(M_i^2-m_i^2)+2\bar{q}_i\sqrt{s}) \right.
\nonumber \\ && \left. \phantom{\frac{2 M}{16 \pi^2} + \frac{\bar{q}_i}{\sqrt{s}}} \hspace*{-0.3cm} - \ln(-s+(M_i^2-m_i^2)+2\bar{q}_i\sqrt{s})\right.
\nonumber \\ && \left. \phantom{\frac{2 M}{16 \pi^2} + \frac{\bar{q}_i}{\sqrt{s}}} \left. \hspace*{-0.3cm} - \ln(-s-(M_i^2-m_i^2) + 2\bar{q}_i\sqrt{s}) \right]
\right\} \nonumber \\
\label{eq:gpropdr}
\end{eqnarray}
where $m_i$ is the mass of the meson in channel $i$. The infinity of integral is canceled
by higher order counter-terms. The subtraction constants $a_i(\mu)$ are real constants, and stand for the finite contribution of such counter-terms. The unknown parameters $a_i(\mu)$ are usually determined by fitting to the data. In this work, we work at the regularization scale $\mu=1200$ MeV and use the following values for the subtraction constants $a_i$~\cite{Inoue:2001ip},
\begin{eqnarray}
&
a_{K^+ \Sigma^-} = -2.8,~~
a_{K^0 \Sigma^0} = -2.8,~~
a_{K^0 \Lambda} = 1.6,
&
\\ \nonumber &
a_{\pi^- p}= 2.0,~~
a_{\pi^0 n}= 2.0,~~
a_{\eta n}= 0.2.
&
\end{eqnarray}
One needs to note that these subtraction constants $a_i(\mu)$ for the channels of $K^0\Lambda$, $\eta n$, $\pi^-p$ and $\pi^0 n$ are positive. With a cutoff $q_{\rm max}$ in the cutoff method, the matrix $G_i$ of Eq.~(\ref{eq:gpropdr}) would imply a subtraction constant $a_i(\mu)$ negative, not positive. The need for values $a_i(\mu)>0$ is an indication that one is including the contribution of missing channels in the scattering amplitude~\cite{Hyodo:2008xr}.
However, in the production process of Fig.~\ref{fig:decaydiagram}(b) (see also Eq.~(\ref{eq:hadronization})), the primary $\Lambda_b\rightarrow J/\psi MB$ is selective to just four channels, with particular weights, which then propagate by means of the $G$ function of the figure. We are not justified to use the $G$ function of scattering to account for channels which would not contribute there. For this reason at this point we shall use the ordinary $\tilde{G}_i(M_{\pi^- p})$ function with a cut off in the following analyses,
\begin{eqnarray}
\label{Loop_integral}
\tilde{G}_i(M_{\pi^-p})&=&\int \frac{d^3q}{{(2\pi)}^3}\frac{M_i}{2\omega_i(q)E_i(q)}\nonumber \\ &&
\times \frac{1}{M_{\pi^-p}-\omega_i(q)-E_i(q)+{\rm i}\epsilon},
\end{eqnarray}
where $M_i$, $E_i$ and $\omega_i$ are the baryon mass, baryon energy and meson energy of the $i$ channel. We regularize by a cut off $|\vec{q}_{\rm max}|=1300$ MeV, but the behaviour of the $J/\psi p$ distribution around the peak is not changed if other values are used.

\subsection{Amplitudes with and without the $P_c(4450)$}
\label{sec:pentaquark}
The next step is to take into account the final state interaction of the primarily produced meson and baryon pairs of Eq.~(\ref{eq:hadron}).
We can write the amplitude $\mathcal{M}(M_{J/\psi p},M_{\pi^- p})$ for $\Lambda^0_b \to J/\psi p \pi^-$, still without the effect of the $P_c(4450)$, as a function of the invariant masses $M_{J/\psi p}$ and $M_{\pi^- p}$,
\begin{eqnarray}
&&\mathcal{M}(M_{J/\psi p}, M_{\pi^- p})\nonumber
\\&=& V_p \left[ h_{\pi^- p} + \sum_{i} h_i \tilde{G}_i(M_{\pi^- p}) \, t_{i,\pi^- p}(M_{\pi^- p}) \right]
\nonumber \\ &=&V_p \left( h_{\pi^- p} + T_{\pi^- p} \right) ,
\label{eqn:halfamplitude}
\end{eqnarray}
where the coefficients $h_i$ are taken from Eq.~(\ref{eq:hadron}),
\begin{eqnarray}
&
h_{K^+ \Sigma^-} = 0,~~
h_{K^0 \Sigma^0} = 0,~~
h_{K^0 \Lambda} = \sqrt{\frac{2}{3}},
&\nonumber
\\  &h_{\pi^- p}= 1,~~
h_{\pi^0 n}= - {1 \over \sqrt2},~~
h_{\eta n}= {1 \over \sqrt3}.
& \label{eq:weight}
\end{eqnarray}
The factor $V_p$ expresses the weak and hadronization strength, and it also contains the overall suppression factor $V_{cb}V^*_{cd}$. We note that the $\tilde{G}_i(M_{\pi^- p})$ function here is calculated by using the cut off method with $|\vec{q}_{\rm max}|=1300$ MeV, as listed in Eq.~(\ref{Loop_integral}).

Now we take into account the $P_c(4450)$ pentaquark contribution, as depicted in Fig.~\ref{fig:decaydiagram}(c). We can see that the $\pi^-p$ production proceeds both at the tree level, as shown in Fig.~\ref{fig:decaydiagram}(a), and through rescattering of other original meson baryon pairs, as depicted in Fig.~\ref{fig:decaydiagram}(b).
In addition, if the pentaquark  signal is a consequence of the excitation of a molecular state, it can be taken into account  in the present approach by allowing for the $J/\psi p$ pair interaction. This is done by means of the
diagram of Fig.~\ref{fig:decaydiagram}(c). Altogether, the $\Lambda^0_b\rightarrow J/\psi p \pi^-$ amplitude $\mathcal{M}(M_{J/\psi p},M_{\pi^- p})$ can be written as,
\begin{eqnarray}
&&\mathcal{M}(M_{J/\psi p}, M_{\pi^- p})\nonumber
\\&=& V_p \left[ h_{\pi^- p} + \sum_{i} h_i \tilde{G}_i(M_{\pi^- p}) \, t_{i,\pi^- p}(M_{\pi^- p}) \right.
\nonumber \\ && \left. \phantom{\sum_{i}}\hspace*{-0.2cm} + h_{\pi^- p} G_{J/\psi p}(M_{J/\psi p})\,t_{J/\psi p, J/\psi p}(M_{J/\psi p})\right] \,
\nonumber \\ &=& V_p \left( h_{\pi^- p} + T_{\pi^- p} + T_{J/\psi p}\right) ,
\label{eqn:fullamplitude}
\end{eqnarray}
where $G_{J/\psi p}$ is the $G$-function representing the loop integral of $J/\psi$ and proton, for which we use the same dimensional regularization as Eq.~(\ref{eq:gpropdr}), but with the regularization scale $\mu=1000$ MeV and the subtraction constant $a_{J/\psi p} = -2.3$~\cite{Wu:2010jy,wumas}. The coherent sum of $T_{J/\psi} p$ in Eq.~(\ref{eqn:fullamplitude}) in $S$-wave holds strictly when
$J/\psi p$ is in total spin $J=1/2$, which is one of the possible spins of the hidden charm states predicted in Ref.~\cite{wumas}. The case for $J=3/2$ is
explicitly done in Ref.~\cite{Lu:2016gev}.
%From this equation, we can clearly see that $V_p$, which contains the suppressed factor $V_{cb}V^*_{cd}$, is indeed an overall factor.

Following the steps of Ref.~\cite{Roca:2015dva}, we have,
\begin{equation}
t_{J/\psi p , J/\psi p}(M_{J/\psi p}) = \frac{g^2_{J/\psi p}}{M_{J/\psi p} - M_{P_c} + i\frac{\Gamma_{P_c}}{2}} \, . \label{eq:amppc}
\end{equation}
where the  three parameters have been fixed in Ref.~\cite{Roca:2015dva}:
\begin{equation}
g_{J/\psi p} = 0.6 \, , \, M_{P_c} = 4449.8 \, {\rm MeV} \, , \, \Gamma_{P_c} = 40 \, {\rm MeV} \, .
\end{equation}

Eq.~(\ref{eqn:fullamplitude}) can be used to calculate the invariant mass distribution of the process $\Lambda^0_b \to J/\psi p \pi^-$:
\begin{align}
\label{eqn:dGammadM}
\frac{d^2\Gamma}{dM^2_{J/\psi p}dM^2_{\pi^- p}}
=\frac{1}{(2\pi)^3}\frac{4M_{\Lambda_b^0} M_p}{32M^3_{\Lambda_b^0}} \left|\mathcal{M}(M_{J/\psi p}, M_{\pi^- p})\right|^2\,,\end{align}
This is a three-body decay and the invariant mass distribution with respect to any of the two invariant masses is evaluated by integrating over the other invariant mass.

Finally, we take into account the $P$-wave and  $D$-wave in $\pi^- p$ scattering to create the Roper $N^*(1440)$ and $N^*(1520)$, respectively, because there are signals of the excitation of both resonances in the $\pi^-p$ mass distribution~\cite{Aaij:2014zoa}. Since the contributions of $P$-wave and $D$-wave add incoherently to the other contributions, we take them into account by means of the substitution below,
\begin{eqnarray}
&&\left|\mathcal{M}(M_{J/\psi p}, M_{\pi^- p})\right|^2  \nonumber \\
& \Rightarrow& \left|\mathcal{M}(M_{J/\psi p}, M_{\pi^- p})\right|^2 + \left|\mathcal{M}^P(M_{\pi^-p})\right|^2 \nonumber \\ && + \left|\mathcal{M}^D(M_{\pi^-p})\right|^2  , \nonumber\\ \label{eq:ampwave}
\end{eqnarray}
with
\begin{eqnarray}
\mathcal{M}^P(M_{\pi^-p})&=& - V^P_p{\rm cos}\theta\frac{\tilde{p}_{\pi^-}}{ (\tilde{p}_{\pi^-})_{\rm ave}}\frac{M_{N^*(1440)}}{\pi}\nonumber\\ &&\times  {\rm Im}\frac{1}{M_{\pi^-p}-M_{N^*(1440)}+i\frac{\Gamma_{N^*(1440)}}{2}}, \label{eq:termP} \\
\mathcal{M}^D(M_{\pi^-p})&=& - V^D_p\frac{3{\rm cos}^2\theta-1}{2}\left( \frac{\tilde{p}_{\pi^-}}{ (\tilde{p}_{\pi^-})_{\rm ave}}\right)^2\frac{M_{N^*(1520)}}{\pi}\nonumber\\ &&\times  {\rm Im}\frac{1}{M_{\pi^-p}-M_{N^*(1520)}+i\frac{\Gamma_{N^*(1520)}}{2}}, \label{eq:termD}
\end{eqnarray}
where $M_{N^*}$ and $\Gamma_{N^*}$ are the mass and width of the Roper $N^*(1440)$ or $N^*(1520)$, respectively, and $V^P_p$ and $V^D_p$ stands for the strength of the $P$- and $D$-wave amplitudes. Both of them are  free parameters independent of $V_p$, but only their ratios $V^P_p/V_p$ and $V^D_p/V_p$ matter up to a global normalization, which will be fitted using the LHCb data~\cite{Aaij:2014zoa}. In Eqs.~(\ref{eq:termP}) and (\ref{eq:termD}), $\tilde{p}_{\pi^-}$ is the $\pi^-$ momentum in the $\pi^-p$ rest frame, and $(\tilde{p}_{\pi^-})_{\rm ave}$ is an average momentum taken for $(\tilde{p}_{\pi^-})_{\rm ave}=(M^{\rm min}_{\pi^-p}+M^{\rm max}_{\pi^-p})/2$, which is put here to have  $V^P_p$ and $V^D_p$  with the same dimensions as $V_p$. The other parameter $\theta$ is the angle between the momentum of $\pi^-$ and $J/\psi$ in the rest frame of the $\pi^-p$ system,
\begin{equation}
{\rm cos}\theta=\frac{1}{2\tilde{p}_{J/\psi}\tilde{p}_{\pi^-}}\left( M^2_{J/\psi p}-M^2_{\Lambda^0_b}-m^2_{\pi^-}+2\tilde{p}^0_{\Lambda^0_b}\tilde{p}^0_{\pi^-} \right) \, ,
\label{eqn:costheta}
\end{equation}
where $\tilde{p}^0_{\Lambda^0_b}$ ($\tilde{p}^0_{\pi^-}$) is the energy of $\Lambda^0_b$ ($\pi^-$)
in the $\pi^- p$ rest frame, and $\tilde{p}_{J/\psi}=\tilde{p}_{\Lambda^0_b}$ ($\tilde{p}_{\pi^-}$)
is the $J/\psi$ ($\pi^-$) momentum in this same frame, where
 $\vec{\tilde{p}}_{\Lambda^0_b}-\vec{\tilde{p}}_{J/\psi}=0$. We give the explicit forms for those variables below,
\begin{eqnarray}
\tilde{p}_{\Lambda^0_b} &=& \frac{\lambda^{1/2}\left(M^2_{\Lambda^0_b}, M^2_{\pi^-p}, m^2_{J/\psi} \right)}{2M_{\pi^-p}} = \tilde{p}_{J/\psi}, \nonumber \\
\tilde{p}^0_{\Lambda^0_b} &=& \sqrt{M^2_{\Lambda^0_b} + \tilde{p}^2_{\Lambda^0_b}}, \nonumber \\
\tilde{p}_{\pi^-} &=& \frac{\lambda^{1/2}\left(M^2_{\pi^-p}, m^2_{\pi}, M^2_p \right)}{2M_{\pi^-p}}, \nonumber \\
\tilde{p}^0_{\pi^-} &=& \frac{M^2_{\pi^-p}+m^2_\pi-M^2_p}{2M_{\pi^-p}},
\end{eqnarray}
where $\lambda(x,y,z)=x^2+y^2+z^2-2xy-2yz-2xz$.

\section{Results and Discussion}
\label{sec:results}
\begin{figure}[!htb]
\centering
  \includegraphics[width=0.45\textwidth]{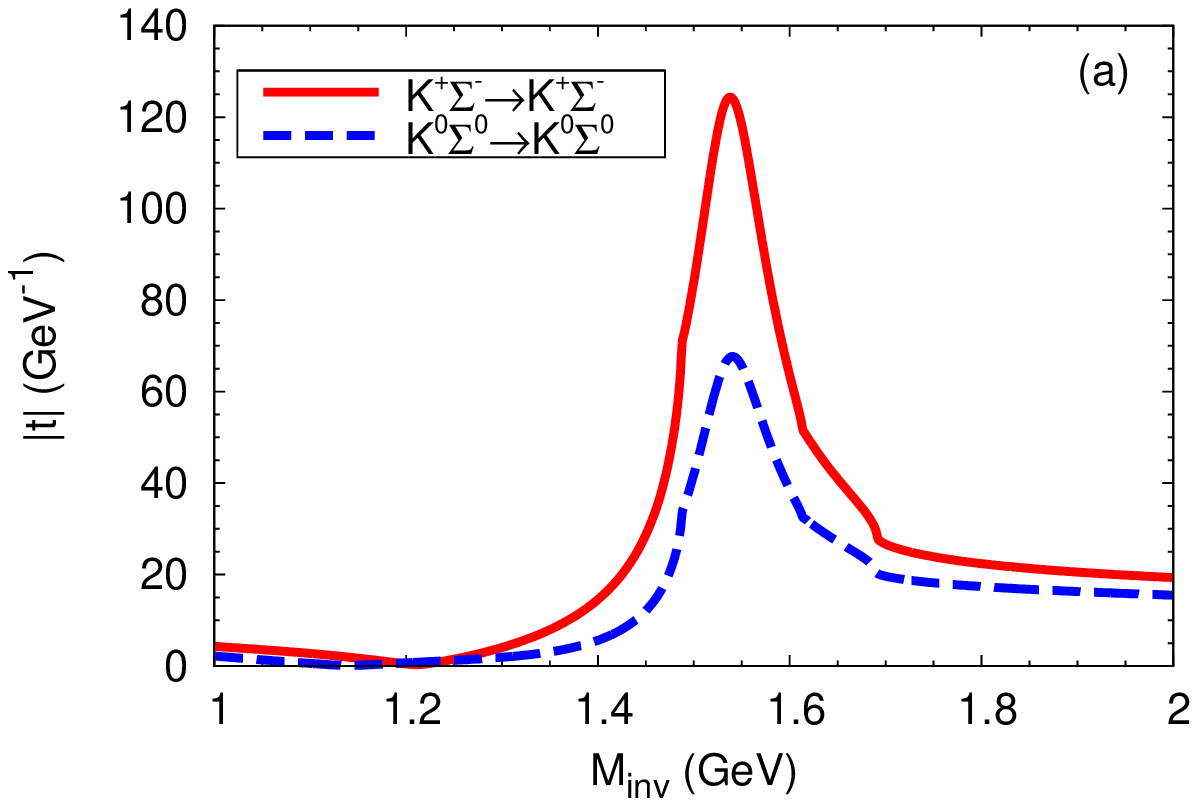}
  \includegraphics[width=0.45\textwidth]{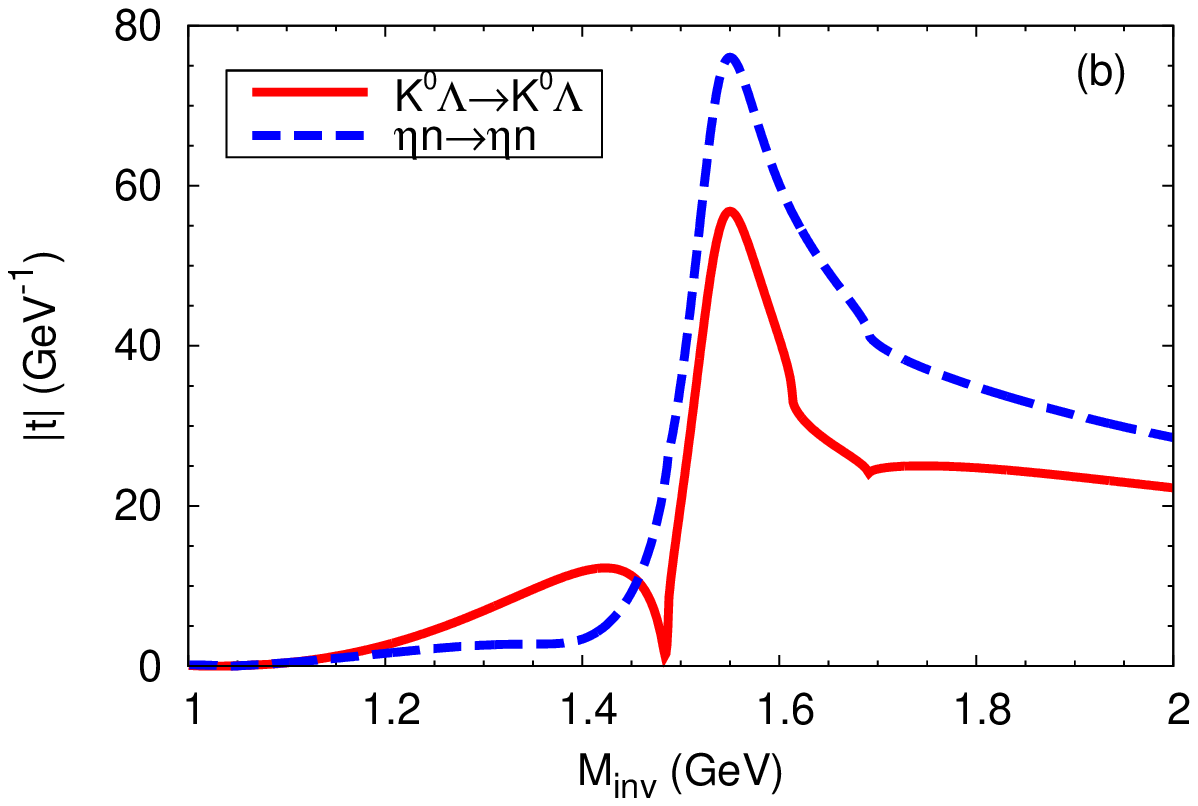}
  \includegraphics[width=0.45\textwidth]{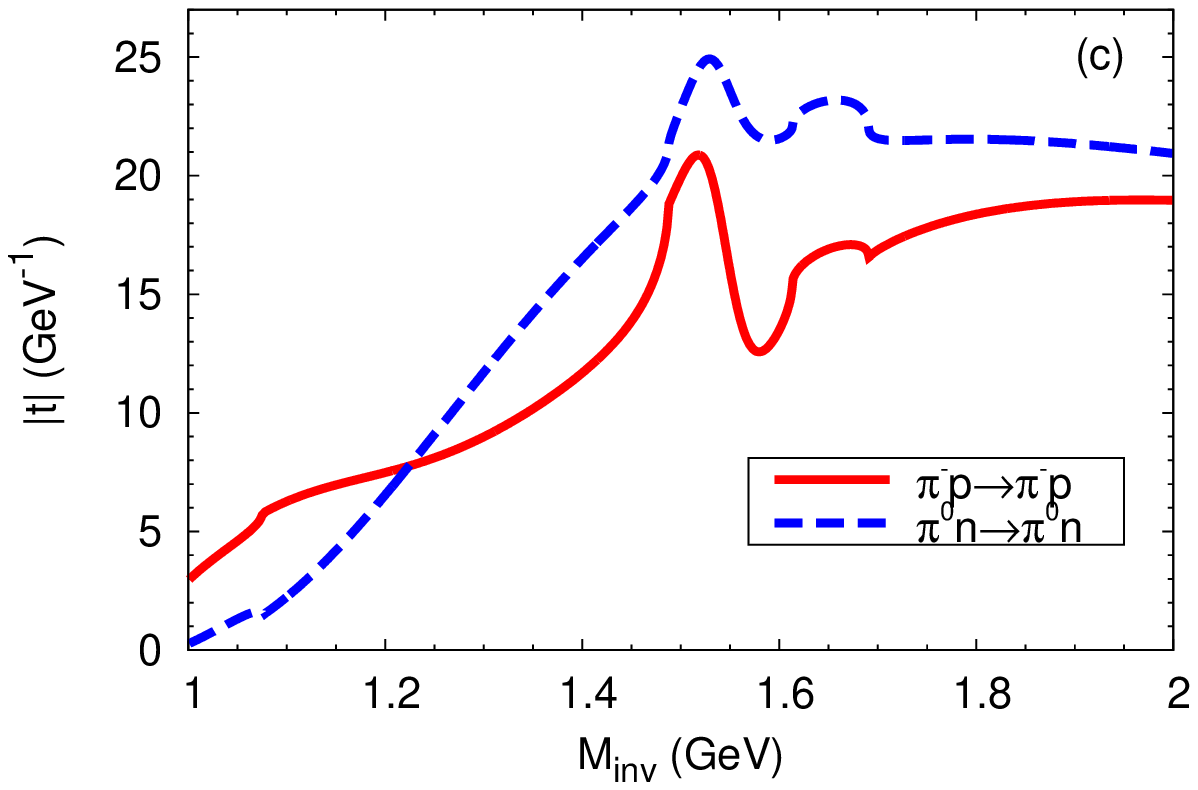}
%\vspace{0.5cm}
\caption{(Color online) Modules of the transition amplitude $|t_{ij}|$ as functions of the invariant mass of the meson-baryon system: (a) $K^+\Sigma^-\rightarrow K^+\Sigma^-$ and $K^0\Sigma^0\rightarrow K^0\Sigma^0$, (b) $K^0\Lambda\rightarrow K^0\Lambda$ and $\eta n\rightarrow \eta n$, (c) $\pi^-p\rightarrow \pi^-p$ and $\pi^0n\rightarrow \pi^0n$.}
  \label{fig:moduleT}
\end{figure}

\begin{figure*}[!htb]
\centering
     \includegraphics[width=0.7\textwidth]{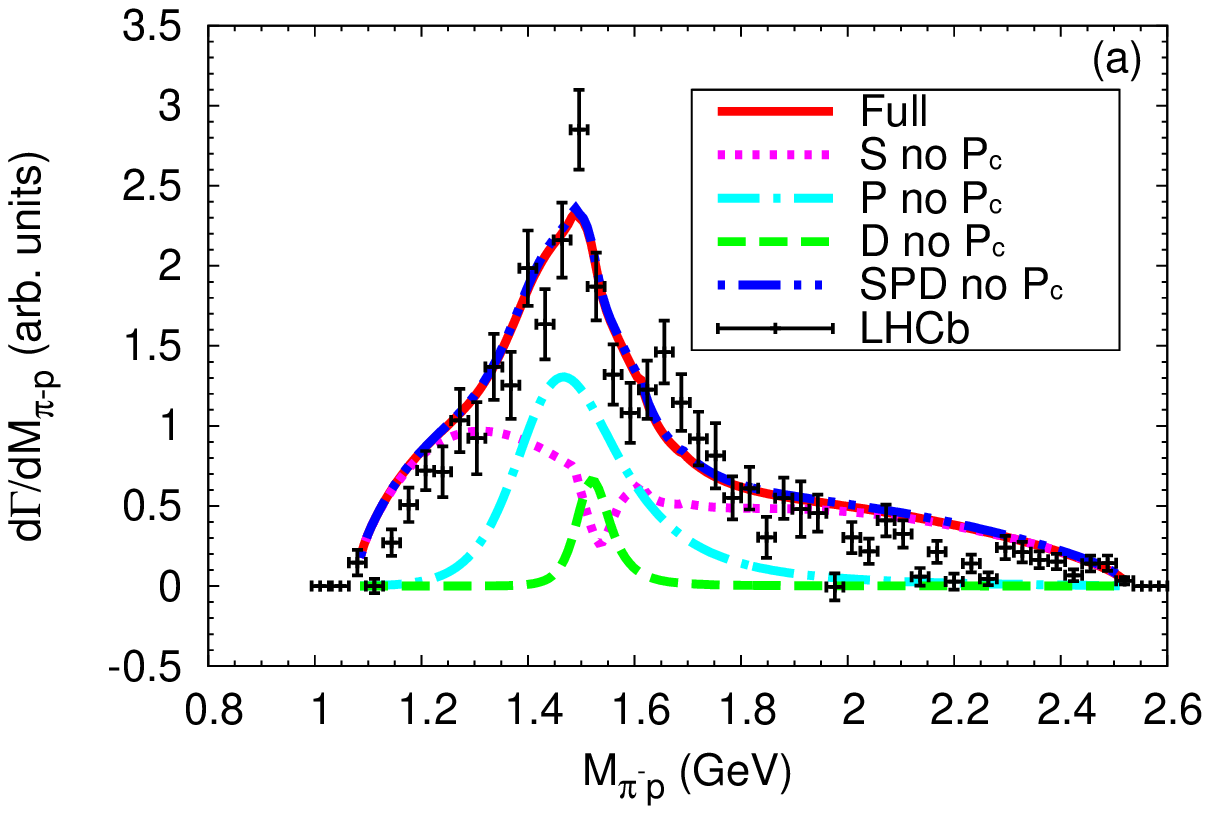}
      \includegraphics[width=0.7\textwidth]{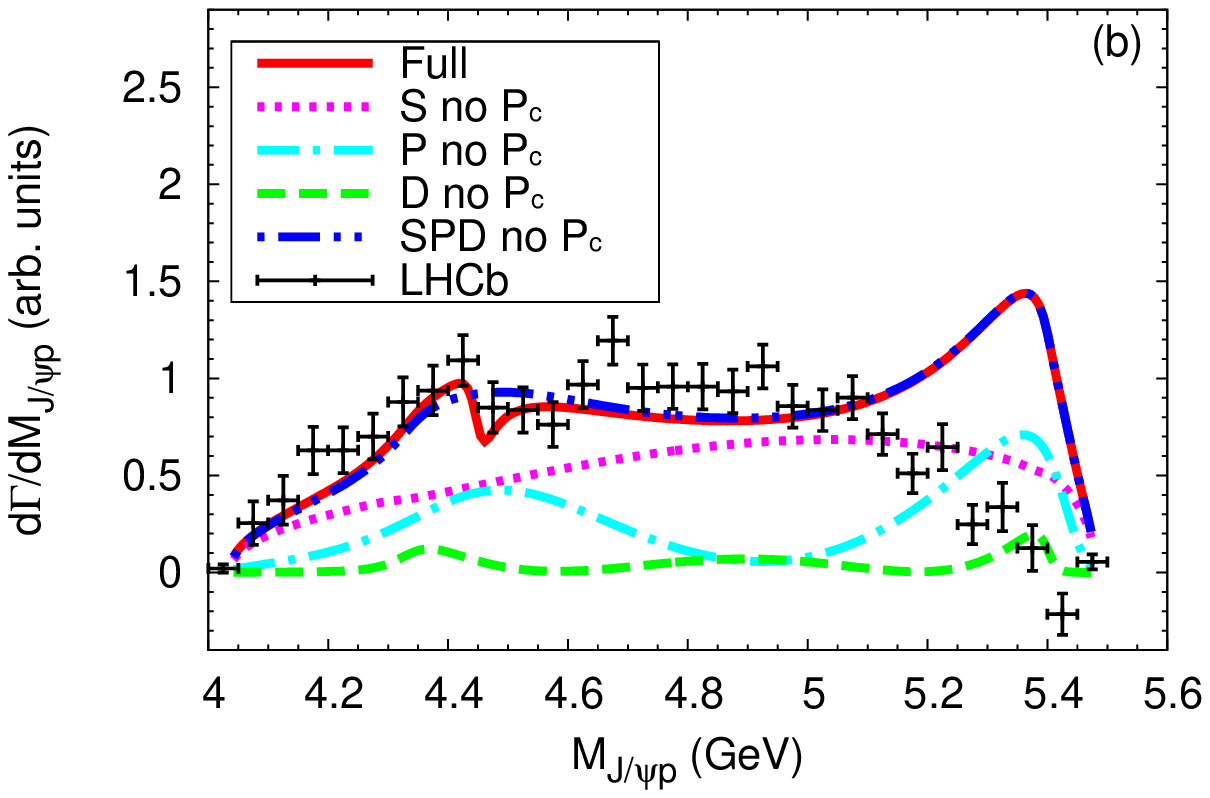}
%\vspace{0.5cm}
\caption{(Color online) (a) The $\pi^- p$ invariant mass distributions and (b) the $J/\psi p$ invariant mass distributions for the $\Lambda^0_b\rightarrow J/\psi p \pi^-$ decay from Ref.~\cite{Aaij:2014zoa}. The magenta dotted line shows the result of the tree level and the $S$-wave $\pi^-p$ interaction [the term $h_{\pi^-p}+T_{\pi^-p}$ of Eq.~(\ref{eqn:fullamplitude})], the cyan dashed-dotted and green dashed lines correspond to the contribution from the $P$- and $D$-wave $\pi^- p$ system alone, and the red solid line stands for the result of our full model. We also show the result without the $J/\psi p$ interaction with the blue dashed-dotted-dotted line. }
  \label{fig:dwidth}
\end{figure*}

In this section, we perform numerical analyses using the formalism described in the previous section. Our results for the process $\Lambda^0_b\rightarrow J/\psi p \pi^-$ will be shown in several figures.

First in Fig.~\ref{fig:moduleT}, we show the modules of the transition amplitudes $|t_{ij}|$ for the coupled channels, $K^+\Sigma^-$, $K^0\Sigma^0$, $K^0\Lambda$, $\eta n $, $\pi^- p$ and $\pi^0 n$ in $I=1/2$ and $S=0$.
Both the $\pi^-p \rightarrow \pi^-p$ and $\pi^0n\rightarrow \pi^0n$ transition amplitudes exhibit a resonance structure around 1535 MeV, which is common to all the unitary chiral approaches~\cite{Kaiser:1995cy,Nacher:1999vg,Nieves:2001wt}. This reaction is the same as in Ref.~\cite{Inoue:2001ip}, where comparison to data is done.

Next we show results for the two invariant mass distributions in the $\Lambda^0_b\rightarrow J/\psi p \pi^-$ process from Ref.~\cite{Aaij:2014zoa}. In Fig.~\ref{fig:dwidth}(a), we show the $\pi^- p$ invariant mass distribution for the $\Lambda^0_b\to J/\psi p \pi^-$ decay. We make a fair fit to the $\pi^-p$ invariant mass distribution by fitting the parameters $V_p$, $V^P_p$ and $V^D_p$. As we can see, the fit demands some $N^*(1440)$ and $N^*(1520)$ contributions to explain the large peak around 1500 MeV, but the $S$-wave contribution accounts mostly for the region of small $\pi^-p$ invariant masses, as one would expect. It also gives contribution at large invariant masses, but this is more uncertain and in any case is not a matter of concern to us, where our aim is to see the consistency of the peak in the $J/\psi p$ mass distribution with the one observed in the $\Lambda^0_b\rightarrow J/\psi pK^-$ reaction. The effect of the $P_c(4450)$ state in this distribution is very small (difference between the ``SPD no Pc'' and ``Full'' curves in the Fig.~\ref{fig:dwidth}(a)).

Next we turn to the $J/\psi p$ distribution which is shown in Fig.~\ref{fig:dwidth}(b). With the choice of $g_{J/\psi p}=0.6$, which is in the range considered acceptable in Ref.~\cite{Roca:2015dva}, we obtain a peak structure around 4400 MeV followed by a dip in this distribution.
The strength of this structure depends on the coupling of the hidden charm state to the $J/\psi p$.
Note, that although the $P_c$ resonance of Eq.~(\ref{eq:amppc}) has a mass of 4450 MeV, at this energy we get here a dip, and the peak-dip structure observed comes from the interference with the rest of the amplitude of Eq.~(\ref{eqn:fullamplitude}). This would justify that the structure seen in this reaction corresponds to the one observed in the $\Lambda^0_b\rightarrow J/\psi p K^-$ reaction even if the peak here is displaced about 40 MeV below the one observed in $\Lambda^0_b\rightarrow J/\psi p K^-$. Note that this behaviour is relatively common in hadron physics. For example, the $f_0(980)$ manifests itself as a clear peak in the $\pi^+\pi^-$ invariant mass of the $J/\psi \rightarrow \phi \pi^+\pi^-$~\cite{Wu:2001vz,Augustin:1988ja} and $B_s\rightarrow J/\psi\pi^+\pi^-$~\cite{Aaij:2011fx} reactions, but shows up as a dip in the $S$-wave $\pi\pi$ scattering amplitude~\cite{Pelaez:2015qba}.

We observe that we do not get a very good agreement for the $J/\psi p$ distribution for invariant masses above 5000 MeV in Fig.~\ref{fig:dwidth}(b). Particularly, we get a peak at the end of the $J/\psi p$ distribution not supported by the experimental data. Our input is only meant to get the region of small and intermediate invariant masses, and in particular to understand the behaviour of the $J/\psi p$ distribution around the peak and consistency with the $\Lambda^0_b\rightarrow J/\psi p K^-$ reaction. We should not worry about the discrepancies in other regions. Because of that we refrain from giving the ratio of the rates of Eqs.~(\ref{brK}) and (\ref{brpi}). A rough estimate of this ratio considering the CKM matrix elements and phase space is given in Ref.~\cite{Aaij:2014zoa}. We should also note that other fits of the quality of Fig.~\ref{fig:dwidth}(a) can be obtained for the $\pi^- p$ distribution changing the various ingredients in it. These changes bring also changes in the $J/\psi p$ distribution, but the peak-dip structure around 4400 MeV is not altered. The main point to stress is that, without the $P_c$ state (contained in our $J/\psi p \to J/\psi p$ amplitude) one obtains a structureless distribution (double dotted-dashed curve in Fig.~\ref{fig:dwidth}(b)), and the inclusion of the $P_c$ state leads to the peak-dip structure (solid line in Fig.~\ref{fig:dwidth}(b)), in agreement with data within errors.

\section{Summary}
\label{sec:summary}

Motivated by the recent LHCb data~\cite{Aaij:2014zoa}, we have studied the $\Lambda_b^0\rightarrow J/\psi p \pi^-$ decay to investigate the hidden-charm pentaquark state within the unitary approach. This model has predicted the existence of two non-strange hidden-charm pentaquark states in the energy region where the
$P_c(4450)$ has been seen. The decay mechanism we employed has been previously adopted to successfully describe the LHCb $\Lambda_b^0\rightarrow J/\psi K^- p$ invariant mass distributions.
Our study showed that  the hidden-charm pentaquark state structure of a peak followed by a dip can be clearly seen on top of the background, which is in agreement with present, low statistics, LHCb data. Given the fact that both the unitary model and the reaction mechanism have been tested in describing the LHCb $\Lambda^0_b$ decay, we look forward to updated experimental results on the $\Lambda_b^0\rightarrow J/\psi p \pi^-$ decay, which can be very helpful to test the existence of the pentaquark states and their nature.

There is an unexpected discovery in the study of this reaction that cannot be let unnoticed. An interesting observation was done in Ref.~\cite{Guo:2015umn} about a possible mechanism that would create the narrow peak in the $\Lambda_b^0\rightarrow J/\psi p K^-$ reaction without having to invoke any new state. It was shown that the triangle diagram of $\Lambda_b\rightarrow \Lambda^*(1890)\chi_{c1}$ followed by $\Lambda^*(1890)\rightarrow K^-p$ decay, which has the intermediate propagators of $\Lambda^*$, $p$, $\chi_{c1}$, develops a singularity when all the three propagators can be put on shell in the loop, and this occurs at the magic invariant mass of 4450 MeV for the $J/\psi p$ distribution. The calculation is done in arbitrary units, because neither the $\Lambda_b\rightarrow \Lambda^*(1890)\chi_{c1}$ nor the  $\Lambda^*(1890)\rightarrow K^-p$ amplitudes are known. However, the fact that the peak in the $\Lambda_b^0\rightarrow J/\psi p \pi^-$ reaction appears about the same energy as in the $\Lambda_b^0\rightarrow J/\psi p K^-$ decay provides a challenge for the
triangular singularity mechanism. One might wonder whether a loop with $N^*p\chi_{c1}$ as intermediate states, with some particular $N^*$, could produce a peak at about the same energy, as seen in the experiment, and with the same relative strength, but this would be a surprising coincidence. This argument reinforces the interpretation of the narrow peak of the $\Lambda_b^0\rightarrow J/\psi p K^-$ as a genuine new exotic baryonic state. Nevertheless, a study along the lines of Ref.~\cite{Guo:2015umn} for the new reaction would be welcome.

{\it Note added:} The study of related reactions along the same lines allows us at this point to provide a broader perspective on what has been done in this paper. The results of Refs.~\cite{Wu:2010jy,wumas} produce a hidden charm pentaquark state that couples mostly to $\bar D^* \Sigma_c$ in $S$-wave, which we hint to correspond to the narrow pentaquark state of Ref.~\cite{Aaij:2015tga}. The interaction used in Refs.~\cite{Wu:2010jy,wumas} borrowed from the local hidden gauge approach~\cite{Bando:1984ej,Bando:1987br,Meissner:1987ge} produces degenerate $\bar D^* \Sigma_c$ states with $J^P = 1/2^-$ and $3/2^-$, but this degeneracy could be broken as shown in Ref.~\cite{Uchino:2015uha}. It is unclear which is the spin of the narrow state in Ref.~\cite{Aaij:2015tga}, because, although several options are preferred, it is also clearly stated that ``other options are less likely'', but this does not mean ruled out. The analysis of the present reaction, being now performed by the LHCb collaboration, will help clarify the issue.

Meanwhile, our present study done in this paper assumes a production vertex $\Lambda_b^0\rightarrow J/\psi p \pi^-$ of $S$-wave. Under these conditions the $J/\psi p$ system can only be in $J=1/2$, because $J=3/2$ with the $\pi^-$ in $S$-wave can not connect with the $1/2^+$ of the original $\Lambda_b^0$. To have $J = 3/2$ for the $J/\psi p$ system one needs a $P$-wave vertex and a different formalism to the present one. This has been done in detail in the work that studies the $\Lambda_b^0\rightarrow J/\psi K \Lambda$ reaction~\cite{Lu:2016gev}, closely related to the present one. We do not wish to repeat that work here and it is sufficient to mention that a signal for the hidden charm state is also seen there, although, depending on the circumstances, it can show as a dip rather than as a peak.

\section{Acknowledgements}

We thank P. Koppenburg for providing us the experimental data points shown in the figures, and T. Gershon, S. Stone, F. K. Guo, W. Wang and T. Burns for useful comments on the paper.
One of us, E. Oset, wishes to acknowledge support from the Chinese Academy of Science (CAS) in the Program of Visiting Professorship for Senior International Scientists (Grant No. 2013T2J0012).
This work is partly supported by the National Natural Science Foundation of China under Grant Nos. 1375024, 11522539, 11505158, and 11475015, the Spanish Ministerio de Economia y Competitividad and European FEDER funds under the contract number
FIS2011-28853-C02-01 and FIS2011-28853-C02-02, and the Generalitat Valenciana in the program Prometeo II-2014/068.  This work is also supported by the China Postdoctoral Science Foundation (No. 2015M582197), and the Postdoctoral Research Sponsorship in Henan Province (No. 2015023). We acknowledge the support of the European Community-Research Infrastructure Integrating Activity Study of Strongly Interacting Matter (acronym HadronPhysics3, Grant Agreement n. 283286) under the Seventh Framework Programme of EU.

\end{document}